\documentclass[conference]{IEEEtran}

\newcommand{\ALT}[2]{#2}
\usepackage{trackchanges}
\addeditor{vg}
\addeditor{dz}
\addeditor{zahra}

\ifCLASSINFOpdf
\else

\fi

\usepackage[ruled]{algorithm2e}
\usepackage[noend]{algpseudocode}
\usepackage{stmaryrd}
\usepackage{marvosym}
\usepackage{booktabs}
\usepackage{enumitem}
 \usepackage{moresize}
\usepackage{dblfloatfix}  
\usepackage{varwidth}
\usepackage{capt-of}
\usepackage{listings}
\usepackage{amsmath}
\usepackage{framed}
\usepackage{tabu}
\usepackage{float}
\usepackage[caption = false]{subfig}
\usepackage{enumitem}
\usepackage{color}
\usepackage{graphicx}
\usepackage{cleveref}
\usepackage{fancybox}
\usepackage[export]{adjustbox}
\usepackage[table]{xcolor}
\definecolor{Gray}{gray}{0.9}
\definecolor{White}{RGB}{255,255,255}
\usepackage{bbold}
\usepackage{bbm}
\usepackage{mathbbol}
\usepackage{multirow}
\usepackage{fix-cm}
\SetKwInput{KwData}{Input}
\usepackage{siunitx,booktabs}
\algnewcommand\algorithmicforeach{\textbf{for each}}
\algdef{S}[FOR]{ForEach}[1]{\algorithmicforeach\ #1\ \algorithmicdo}
\usepackage{xcolor}
\def\headline#1{\hbox to \hsize{\hrulefill\quad\lower.3em\hbox{#1}\quad\hrulefill}}
\usepackage{setspace}
\usepackage{graphicx}
\usepackage{colortbl}
\usepackage{array}
\usepackage{pifont}
\usepackage {rotating}
\usepackage{mwe}
\usepackage{pbox}
\usepackage{enumitem}
\let\oldding\ding
\renewcommand{\ding}[2][1]{\scalebox{#1}{\oldding{#2}}}
\setlist[description]{leftmargin=\parindent,labelindent=\parindent}

\newcommand{\Csharp}{%
  {\settoheight{\dimen0}{C}C\kern-.05em \resizebox{!}{\dimen0}{\raisebox{\depth}{\#}}}}

 \usepackage{enumitem}
\usepackage{xspace}
\usepackage{xparse}
\DeclareDocumentCommand\newstep{o}{%
\item\IfNoValueTF{#1}{}{#1 \textendash\xspace}}

\newlist{steps}{enumerate}{1}
\setlist[steps]{label=\textit{Step \arabic*:},leftmargin=*}
 \usepackage[style=ieee, backend=bibtex, defernumbers=true]{biblatex}

\definecolor{orange}{RGB}{0,32,96}
\definecolor{g}{RGB}{50,50,50}
\definecolor{brightmaroon}{rgb}{0.76, 0.13, 0.28}
\definecolor{byzantine}{rgb}{0.74, 0.2, 0.64}
\definecolor{chromeyellow}{rgb}{1.0, 0.65, 0.0}
\definecolor{applegreen}{rgb}{0.55, 0.5, 0.0}
\definecolor{cadetgrey}{rgb}{0.57, 0.64, 0.69}

\usepackage{capt-of}

\PassOptionsToPackage{gray}{xcolor}
\usepackage{tikz}

\usepackage{tcolorbox}

\begin{document}
\title{ELICA: An Automated Tool for Dynamic Extraction of Requirements Relevant Information}
\author{
    \IEEEauthorblockN{Zahra Shakeri Hossein Abad\IEEEauthorrefmark{1}, Vincenzo Gervasi\IEEEauthorrefmark{2}, Didar Zowghi\IEEEauthorrefmark{3}, Ken Barker\IEEEauthorrefmark{1}}
    \IEEEauthorblockA{\IEEEauthorrefmark{1} Department of Computer Science, University of Calgary, Calgary, Canada, 
\{zshakeri, kbarker\}@ucalgary.ca}
    \IEEEauthorblockA{\IEEEauthorrefmark{2}Department of Computer Science, University of Pisa, Italy, gervasi@di.unipi.it}
    \IEEEauthorblockA{\IEEEauthorrefmark{3}School of Software, University of Technology Sydney, Australia, didar.zowghi@uts.edu.au}
}



%



\maketitle
\makeatletter
\def\ps@IEEEtitlepagestyle{%
  \def\@oddfoot{\mycopyrightnotice}%
  \def\@evenfoot{}%
}
\def\mycopyrightnotice{%
  {\hfill \footnotesize [Preprint version]  2018 IEEE 26$^{\text{th}}$ International Requirements Engineering  Conference Workshops (REW'18) \hfill}
}

\begin{abstract}
Requirements elicitation requires extensive knowledge and deep understanding of the problem domain where the final system will be situated. However, in many software development projects, analysts are required to elicit the requirements from an unfamiliar domain, which often causes communication barriers between analysts and stakeholders. In this paper, we propose a requirements ELICitation Aid tool (ELICA) to help analysts better understand the target application domain by dynamic extraction and labeling of requirements-relevant knowledge. To extract the relevant terms, we leverage the flexibility and power of Weighted Finite State Transducers (WFSTs) in dynamic modeling of natural language processing tasks. In addition to the information conveyed through text, ELICA captures and processes non-linguistic information about the intention of speakers such as their confidence level, analytical tone, and emotions. The extracted information is made available to the analysts as a set of labeled snippets with highlighted relevant terms which can also be exported as an artifact of the Requirements Engineering (RE) process. The application and usefulness of ELICA are demonstrated through a case study. This study shows how pre-existing relevant information about the application domain and the information captured during an elicitation meeting, such as the conversation and stakeholders' intentions, can be captured and used to support analysts achieving their tasks.

  \end{abstract}


\IEEEpeerreviewmaketitle

\section{Introduction and Motivation}
Software development success is highly contingent on the accuracy and relevance of requirements gathered from domain experts, users, and other stakeholders \cite{Elic1}. However, the risk of failing to capture requirements correctly and completely is always present. 
\ALT{\annote[vg]{In May 2017, more than 1.25 million Dodge Ram pickup trucks were recalled worldwide due to a software glitch which can incorrectly detect that a sensor has failed\footnote{https://www.carprousa.com/1-25-million-dodge-rams-recalled-airbag-seatbelt-failure}. In this failure, the software did exactly what it was supposed to do. The reason it failed is that it was programmed to do a wrong thing. The completeness and correctness of captured requirements at the elicitation stage depend in large part on effective communication between analysts and stakeholders \protect\cite{MisInfo}.}{This could be safely removed to recoup space; people at RE know about these risks.}} There are many elicitation techniques available to help analysts extract requirements from different sources, such as interviews, questionnaires, introspection \cite{Elic1}, and observation \cite{Irit}. Appan and Brown \cite{MisInfo} conducted a series of experiments on the issue of memory recall and lack of domain knowledge during requirements elicitation. They noted that introduction of misinformation by analysts during a requirements elicitation meeting reduces the accuracy of requirements provided by stakeholders. This makes interviews more vulnerable to the misinformation effect than other elicitation techniques such as the questionnaire. Moreover, during an interview, analysts have primary control over the requirements elicitation process and their domain knowledge in the context of the system at hand and their ability to recall relevant information may impact the completeness and correctness of the elicited requirements. 

Over the years, researchers have contributed valuable techniques and tools to either manage the information gathered during the elicitation process or to provide cognitive support to analysts \cite{Oli,Tool1,Elicito,Tool2,Tool3,REETA}. However, practitioners do not widely use these tools which implies research has yet to fully address analysts' needs during the elicitation process to better understand the domain of the target application and uncover users' requirements \cite{Elic1, Cog}.
In this paper we present a {tool-supported approach} (called ELICA---ELICitation Aid) that provides a suite of innovative automated techniques by which interaction between an analyst and one or more stakeholders is processed in real-time (we include spoken interaction, via a third-party speech-transcription utility, and written {\em in-context} interaction, e.g. in chat or quick-turn emails). We propose to use generative models, based on Weighted Finite State Transducers {\small (WFSTs)} \cite{Mohri} and statistical Language Models {\small (LMs)} to extract requirements-relevant knowledge from the existing documents. The flexibility of WFSTs in modeling {\em variable-length} textual snippets allows the easy integration of input texts with our proposed generative model. 

 Regardless of the elicitation technique employed, in addition to the information conveyed through text, ELICA captures and processes non-linguistic information about the intention of interviewees such as their confidence level, analytical tone, and emotions. This additional information will be used as a complementary source of data in the interpretation of the extracted information and assists analysts in understanding better the tone and aspiration of stakeholders. Furthermore, ELICA provides visual aids for analysts by highlighting the most relevant terms in the extracted snippets as well as visualizing intentions. All the information generated during the elicitation process is made available to the stakeholders as an additional artifact. This enables all stakeholders to obtain details about the information generated and transferred during the elicitation process.

\section{Related Work}
\label{sec:RW}
In this section, we briefly survey the related work on tool support for elicitation in two categories relevant to the scope of this paper: (1) collaborative media-based tools which support collaborative elicitation and augment communication and discussion among analysts and stakeholders, and (2) natural language (NL) tools that process requirements specification documents and text. 

\subsection{Collaborative Media-based Tools}
There is a growing body of research investigating the application of multi-media for dynamic elicitation of requirements \cite{Tool0, Oli, Tool1, Collab}. In an early work, Kaiya et al. \cite{Tool0} developed a tool to record and (manually) structure the minutes discussed during elicitation meetings. They used a set of pre-defined keywords to extract the main topics of discussions and used chronological and Is-A relationships to structure these topics. 
Gall et al. \cite{Tool1} proposed a framework which uses video to record requirement elicitation meetings and automatically extract important statements raised by stakeholders. Karras et al. \cite{Oli} proposed a video analysis tool which combines textual minutes with their corresponding part of the video recorded during a requirements elicitation meeting. The highlighted relevant sections of the video which contain both verbal and non-verbal information along with the attached summarized notes for each section can be used as a source of information for requirements elicitation task. 

To involve clients actively in the process of elicitation, P\'{e}rez and Valderas \cite{Clients} developed a tool which allows end-users to use interactive visualization to describe the main characteristics of the pervasive system. Likewise, to engage and stimulate users to the elicitation process, Duarte et al. \cite{Client2} proposed a web-based collaborative environment for requirements elicitation, with both requirements and social visualization support. Coulin and Zowghi \cite{Collab} proposed a situational collaborative tool called MUSTER which aims to enable multiple stakeholders to work collaboratively with each other and with analysts within an elicitation workshop.

\subsection{Natural Language Tools}
The application of ontologies is one of the popular directions in NL-oriented tools to support exploratory elicitation. Kitamura et al. \cite{Ontology3} proposed an ontology-based tool, using Prolog, which utilizes the domain ontology of a system as domain knowledge to help analysts evolve requirements systematically by providing more information about the semantic aspects of requirements. Elicito \cite{Elicito}, proposed by Balushi et al., applies quality requirements ontologies to provide a knowledge repository of quality requirements (i.e. NFRs) which can be used as a memory aid during elicitation interviews. This repository helps analysts structure interviews and guide them with respect to the important quality aspects relating to the system. To improve the communication activity during the RE process, Valderas and  Pelechano \cite{Tool2} proposed a tool based on requirements ontologies which provides customers with an intuitive interface to describe their needs. Then, the ontology-based descriptions of clients' needs will be transformed into a textual requirements description which will be used for further analysis by analysts. In a similar vein, Farfeleder et al. \cite{Tool3} presented a tool that uses relations and axioms of a domain ontology to support analysts by providing semantic guidelines (a list of suggestions) during the requirements elicitation task. 

With regard to the application of text mining techniques in elicitation tools, Sampaio et al. \cite{Sampaio} presented a tool, called EA-minder, that utilizes NLP features from
WMATRIX to provide an automated support for analysts by identifying the important parts of the input document. Lian et al. \cite{Mona} proposed a Mining Requirements Knowledge (MaRK) tool to reduce human effort during the elicitation process. MaRK provides a semi-automated support for analysts tasked with extracting requirements knowledge in any phase of a software project. In a recent work, Noaeen et al. \cite{REETA} developed a requirements elicitation tool to capture requirements of transportation engineering management systems (called RETTA). This tool leverages public crowd-sourcing paradigm (by crawling social networks and using open traffic sensors' data) as a means to gather richer data repository for text analysis algorithms (i.e. Na\"{i}ve Bayes and LDA \cite{Data17}).

While the existing tools offer a variety of features to assist analysts during the elicitation process, they typically apply pre-learned information or require domain experts' involvement to extract requirements-relevant information. 

\section{Problem Formulation and Preliminaries}

\subsection{Problem Formulation}
\label{sec:Formula}
Given a {\em document repository} ${\cal D}=\{d_1, \ldots, d_n\}$, with each $d_i$ representing a document, and a {\em source stream} (a real-time transcription of an ongoing interview), ${\cal S}= \langle{s_1, s_2, \ldots s_m}\rangle$ with each $s_i=(a_i, b_i)$ representing an {\em exchange} (where  without loss of generality $a_i$ indicates text from the analyst, $b_i$ indicates text from the stakeholder), our tool ELICA addresses the following scenarios:

\begin{figure}
\centering
\includegraphics[scale=.4]{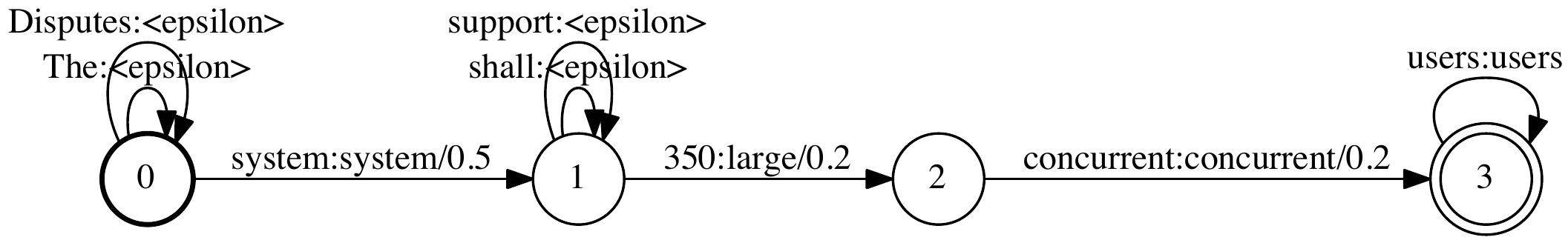}
\caption{\scriptsize An example of using WFST to model natural language text. This WFST replaces English and contextual stop-words with empty string $\varepsilon$. It also replaces numbers with a contextual term (contextual replacements are out of the scope of this paper). If we define the weight as the cost of each transition, the lower the weight, the higher the probability of the term in a context. In this example, relevant terms {\em large} and {\em concurrent} are receiving lower weights. }
\label{fig:WFST}
\end{figure}
\begin{figure}
\centering
\subfloat [$T_1$]{\includegraphics[scale=0.5]{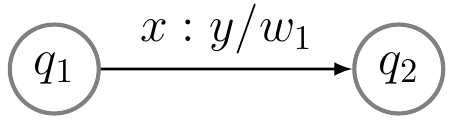}}\hfill
\subfloat [$T_2$]{\includegraphics[scale=0.5]{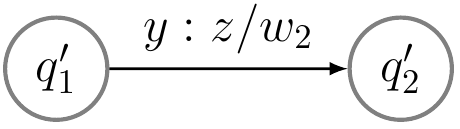}}\hfill
\subfloat [$T_1 \circ T_2$]{\includegraphics[scale=0.5]{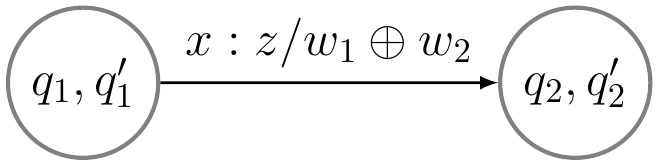}}
\caption{The composition ($\circ$) operation for detecting substrings (subsequences) of interest with transition rule: $(q_1, x, y, w_1, q_2) \circ (q_1', y, z, w_2, q_2') \Rightarrow$ $((q_1, q_1'), x, z, w_1\oplus w_2, (q_2, q_2'))$}
\label{fig:Composite}
\end{figure} 



\begin{figure*}[!b]
\centering

\includegraphics[scale=0.7]{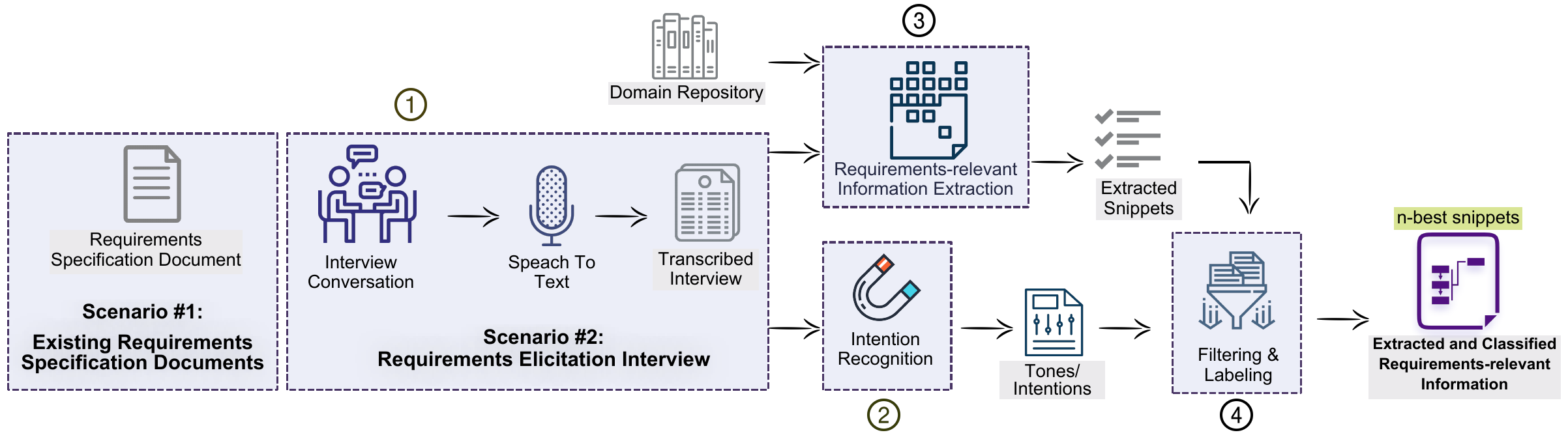}

\caption{The overall process of extracting requirements relevant knowledge implemented in ELICA.}
\label{fig:Process}
\end{figure*}

\begin{description}
\item [Scenario \#1:] An analyst is assigned to an ongoing project and she needs to become familiar with a possibly large amount of documentation that has already been produced. ELICA uses the existing problem domain documents in ${\cal D}$ to select textual snippets that are most relevant to the part of the system under investigation. In this scenario, $s_i$ presents a textual snippet of the existing documentation. 
\item [Scenario \#2:] An analyst is assigned to work on the requirements for a project from an unfamiliar domain. To support the analyst during the elicitation meeting, ELICA selects inside each $d_i$ those textual snippets that are most {\em relevant} for {the most recent part} of the conversation happening in ${\cal S}$. 
\end{description}
Selection of requirements-relevant snippets (${\cal R}=\{r_1,  \ldots r_v\}$) will constitute selecting textual spans which are most likely to be relevant given surrounding context. The context of an occurrence is defined by substrings or subsequences (non-contiguous terms) surrounding it. Each relevant snippet ${r_j}$ contains a set of requirements-relevant terms $\{t_1, t_2, \ldots, t_k\} \in {\cal T}$. 
{Fig.~\ref{fig:Process}} illustrates both scenarios. After identifying ${\cal D}$ and $s_i$, ELICA follows the same process for extracting requirements-relevant knowledge (i.e. most relevant snippet(s)). Given ${\cal D}$ and  ${\cal S}$ as inputs, the output of our technique will be a tuple $\langle {r}, \{t_{1}, t_{2}, \ldots,  t_{z}\} \rangle $, where $z$ is defined by analyst and represents the maximum number of relevant terms that should be highlighted in each extracted snippet.

Moreover, ELICA can be applied in a situation where an analyst is working on a requirements exploration task (e.g. elicitation meeting, discussion with other team members, ...) and needs to switch this task to address an incoming task. To assist analysts to manage the issues of memory recall \cite{RE17, EASE18} after resuming the elicitation task, ELICA uses ${\cal S}$, produced during the exploration task, as well as the existing problem domain documents (i.e. ${\cal D}$) and provides relevant information needed to resume the switched task.


\subsection{Preliminaries}
We briefly describe some of the main theoretical and algorithmic aspects of WFST machines.

{\setlength\parindent{5pt}
\paragraph{{Weighted Transducers}} 
A Finite State Transducer (FST) is a finite automaton in which a successful path through the initial state to a final state represents a mapping from an input sequence (i.e. characters, words) to an output string \cite{Mohri}.
A weighted transducer is an FST that adds a weight to each transition, in addition to the input and output strings. This weight may encode probabilities, priority, or any other quantities assigned to alternative, uncertain transitions.}

Figure \ref{fig:WFST} gives a simple example of a WFST to model sample requirement ``The Disputes system shall support 350 concurrent users". The input and output labels $x$ and $y$, and weight $w$ of a transition are presented on transition arcs by {\em x:y/$w$}. 
{\setlength\parindent{5pt}
\paragraph{{Composition of WFSTs}} 
WFSTs can be composed by a general operation for tying two or more WFSTs together to create a pipeline which can be used to represent statistical models of both generative and discriminative models (e.g. Language Models (LM) and SVMs). As illustrated in Figure \ref{fig:Composite}(a-c), given two WFSTs $T_1$ and $T_2$ such that the output alphabet of $T_1$ coincides with the input alphabet of $T_2$, composition feeds the output of $T_1$ into the input of $T_2$ \cite{Mohri}. Substring $y$ denotes the substring of interest appearing in weighted automata.
}


\section{Proposed Approach}
\label{sec:App}

In this section, we discuss the intuition behind using \emph{contextual lexical association} as well as the technical description of our proposed approach for dynamic control of context-dependency in existing documents. More precisely, the proposed extraction and labeling process is depicted in Figure \ref{fig:Process}. 
More application-minded readers can consult Section \ref{sec:Features} first to get a non-technical summary of the extraction scenario and the state of the field, and then read Sections \ref{sec:PropExtraction} to obtain more technical details. Please note that while ELICA supports the classification of the extracted requirements-relevant information, the classification task is a separate contribution and out of the scope of this work.

\subsection{Rationale}
Recall from Section \ref{sec:RW}, most existing work on the extraction of relevant terms from NL text in the context of RE use basic document features such as terms frequency and length, document length and the existence of a term in a repository. Using these features, relevant terms stay independent of other content-carrying terms in the document which contributes to overlooking the context surrounding terms when measuring their relevance. 
To make the discussion clearer, given the same repository ${\cal D}$, let us consider two source streams ${\cal S}_1$ and ${\cal S}_2$, with each of them being two sentences as follow~\footnote{For the purpose of simplicity, we only list here the stems of content words from a hypothetical exchange; stop words and interlocutory utterances are removed.}: \\
\vspace{-3mm}

${\cal S}_1: s_{11}=\{ \text{\small update location predetermined interval page}\}$

$\textcolor{white}{{\cal S}_1}: s_{12}=\{${\small generate alerts deviate assigned route page}$\}$\\
\vspace{-2mm}

${\cal S}_2: s_{21}=\{ \text{\small page facility upload photograph traffic}\}$

$\textcolor{white}{{\cal S}_2}: s_{22}=\{${\small page display important contact district state}$\}$\\

\vspace{-2mm}
The first source stream (${\cal S}_1$) is describing general requirements of a system (e.g. updating specific locations to a page, and generating alerts for deviations from predefined routes on a page). The term ``page'' is not a content carrying term in this context. On the other hand, ``page'' is a relevant term in ${\cal S}_2$ and two sentences $s_{21}$ and $s_{22}$ talk about requirements of a \emph{page}. Using any of the existing traditional weighting methods (e.g. TF-IDF) which cannot distinguish between these terms (as they do not take the relevance of the document terms/expressions into account), ``page'' will be given approximately the same weigh in both documents ${\cal S}_1$ and ${\cal S}_2$.
{This is a weakness shared by all {\em bag of words} approaches.}
To address this problem, in this paper, we use Weighted Finite State Transducers (WFSTs) to dynamically measure the {\em contextual lexical association} between documents terms, which quantitatively determine the strength of association between two or more words (or terms) based on their co(occurrence) in a corpus \cite{Lex2} and will assign different weights to terms depending on the context they occur in.

\subsection{Overview of Extracting Process}
\label{sec:PropExtraction}
As depicted in Fig.~\ref{fig:Process}, our proposed process for automating the dynamic extraction of requirements-relevant information involves three steps (Step  \ding{195} details the classification task which is out of the scope of this paper):

{\bf Steps \ding{192}, \ding{193} [Data Preparation and Intention Recognition]}: In this step, we form the source stream ${\cal S}$ which will be updated during the elicitation process. In scenario \#1, we obtain ${\cal S}$ from the requirements specification documents provided by the client. in scenario \#2, the latest window of ${\cal S}$ is recorded using a speech-to-text processing engine (see Section \ref{sec:Features}). 
After preprocessing of both ${\cal D}$, and $s_i \in {\cal S}$ for stop word removal and word stemming, $s_i$ will be used as an input to the intention recognition in this step. The main three tones in either $a_i$ or $b_i \in s_i$ which will be identified and used in this step are (from \cite{IBM}):

\begin{itemize}
\item {\em Confident:} A confident tone indicates the $s_i$'s degree of certainty in the ongoing discussion.  
\item {\em Analytical:} An analytical tone indicates the $s_i$'s reasoning and analytical attitude about concepts. 
\item {\em Tentative:} A tentative tone indicates the $s_i$'s degree of inhibition. A tentative $s_i$ can be perceived as a questionable or doubtful information.
\end{itemize}


 {\bf Step \ding{194} [Extraction]}:  To calculate the contextual co-occurrence knowledge (i.e. contextual lexical association) between $s_i$ and ${\cal D}$, our proposed technique builds a static language model for ${\cal D}$ using WFSTs, then recomputes a language model for the most recent ``window'' of ${\cal S}$ (i.e. $s_i$) on every addition to ${\cal S} $. To extract the relevant snippets from ${\cal D}$, we measure the {\em contextual lexical association} \cite{Lex} between $s_i$ and ${\cal D}$. 
To consider all possible variations of a context and to keep our language model from assigning zero to unseen contexts ($n$-grams), we use {\bf $n$-gram hierarchy}. In a nutshell, this approach takes the view that, {sometimes, using \underline{less} context is a good thing and helps to generalize the context of the $n$-gram model.} To implement the $n$-gram hierarchy approach we use and evaluate the following techniques:

\begin{enumerate}
\item {\em Backoff:} Using this technique, if the required $n$-gram has zero counts, we approximate its probability by backing off to $(n-1)$-gram. We continue backing off to a lower-order $n$-gram until we find a term that has a non-zero count. In this paper, we apply and evaluate two backoff methods Katz \cite{Katz} and Witten-bell \cite{Witten}.
\item {\em Interpolation:} Regardless of the frequency of different order $n$-grams, by applying this approach we mix the probability of all the $n$-gram sequences. In other words, we shave off a bit of probability mass from some more frequent terms and give it to the contexts that have never occurred in a corpus \cite{Book}. Kneser-ney \cite{Kneser} and Absolute \cite{Absolute} are the interpolation techniques we use in this paper to implement hierarchical $n$-gram. 
\end{enumerate}
\begin{figure*}
\centering
\subfloat[Perplexity measure]{\includegraphics[scale=0.57]{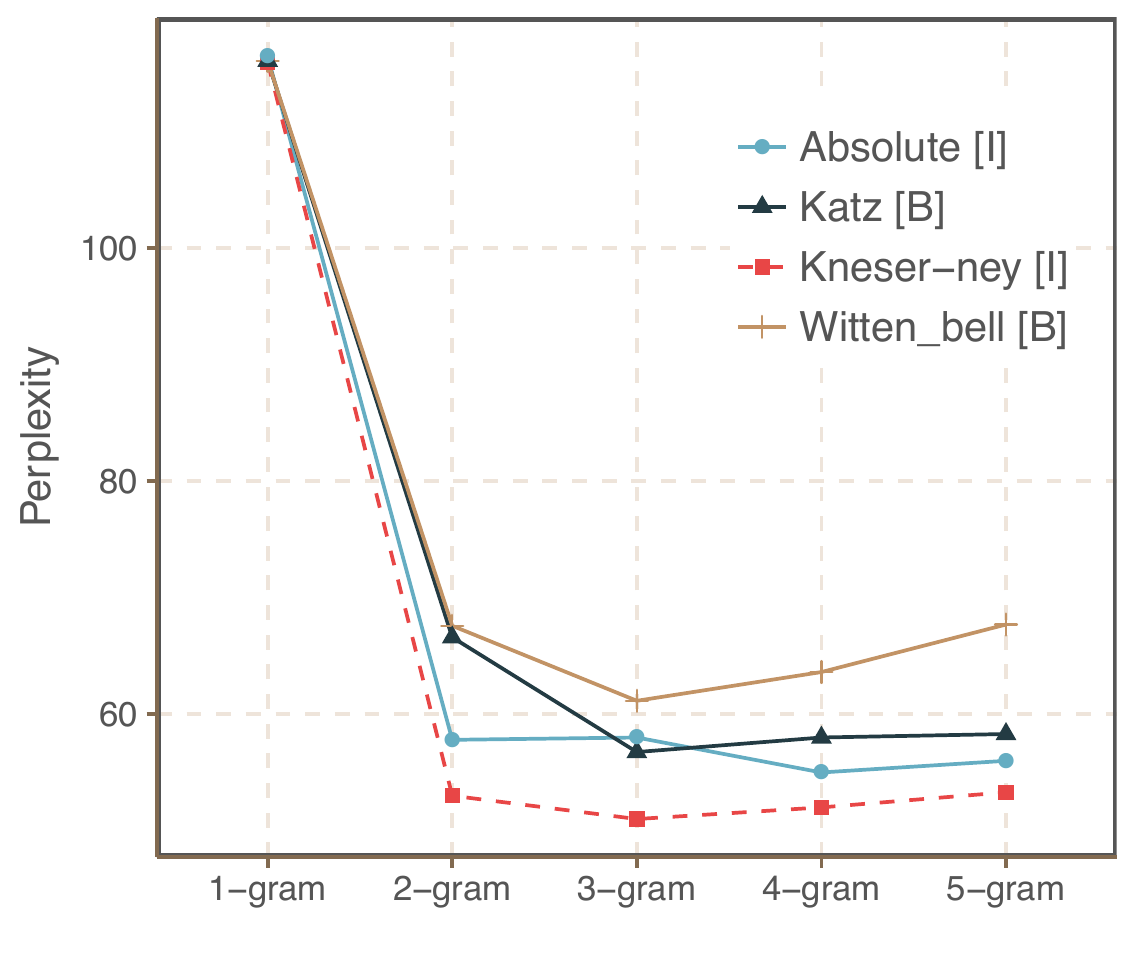}} \hfill
\subfloat [Extracted relevant terms]{\includegraphics[scale=0.43]{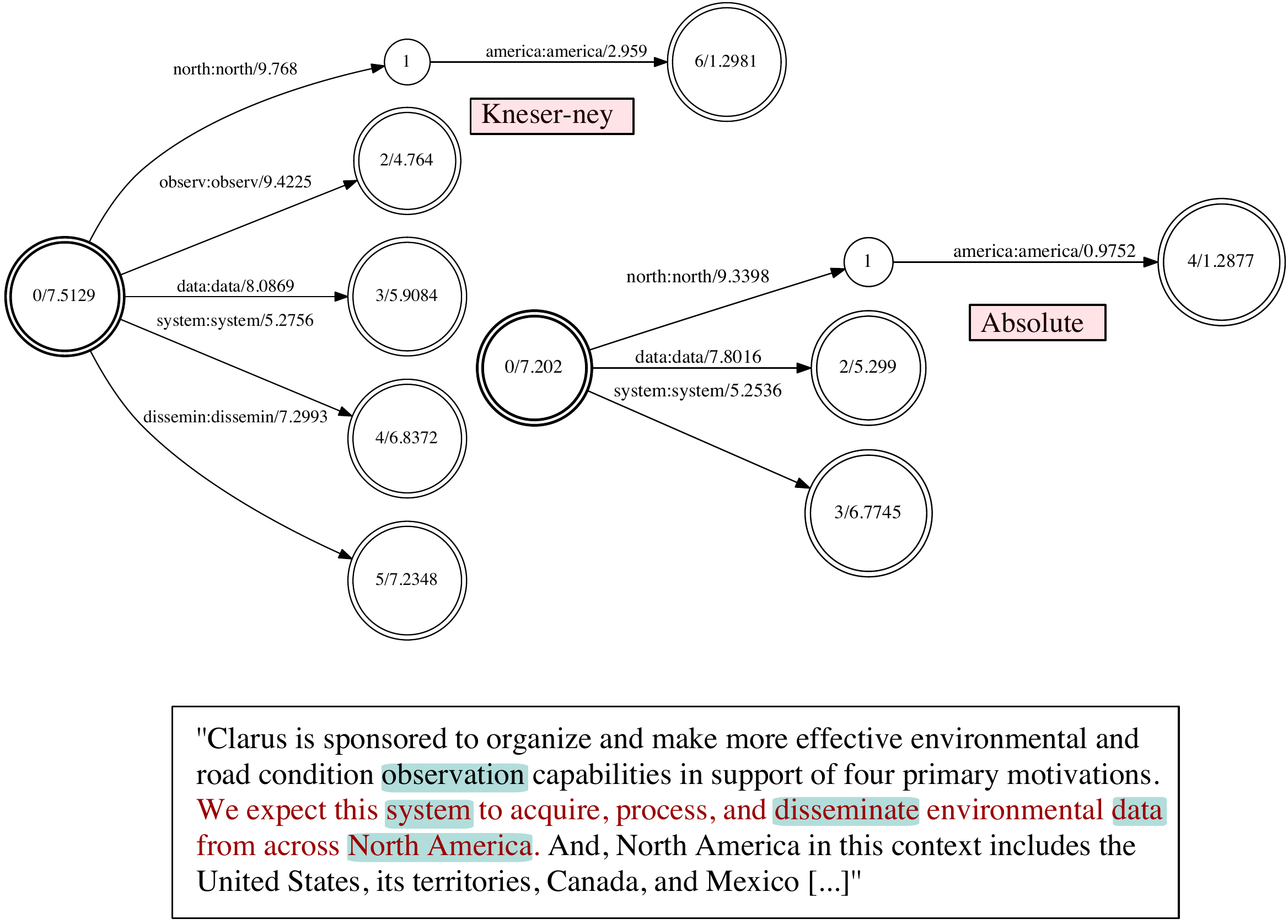}}
\vspace{-2mm}
\caption{An example of the output of the perplexity check and extraction processes. (a)  Perplexity measure for all combinations $\langle m_j, n\rangle$. {[}I{]}: Interpolation, {[}B{]}: Backoff. (b) This example shows the performance of two discounting methods {\em Kneser-ney} and {\em Absolute} in extracting requirements-relevant terms.This confirms the results of the perplexity measure which shows LM$_ {\langle \text{Kneser-ney}, 3\rangle}$ better predicts the most recent ``window'' of ${\cal S}$ (the Absolute method failed to detect ``North America" and ``disseminate'' terms).} 

\label{fig:perpel}

\end{figure*} 
To implement the hierarchical language models, the order of language models ($n$ in $n$-grams), and the details of the structure of $n$-grams (hierarchical $n$-grams) are identified using {\em perplexity measure} \cite{Perpel}. The perplexity of a language model on a test set (which, in this context, is the transcribed interview or a new specification text) is defined as the inverse $n$-gram probability of the source stream data, normalized by the number of words. The lower the perplexity, the higher the ability of the language model in predicting the incoming text.
After transforming ${\cal D}$ to an archive of WFSTs, we use perplexity measure to obtain the most probable LM for each pair of $\langle s_i, {\cal D}\rangle$. In particular, we build 20 static LMs for ${\cal D}$ and index each model with $\langle m_j, n\rangle$, where $m_j \in \{${\small Katz, Witten-bell, Absolute, Kneser-ney}$\}$ and denotes the discounting method, and $n\in\{1, 2, 3, 4, 5\}$ the order of each LM (as an example see Figure \ref{fig:perpel}a). LM$_ {\langle m_j, n\rangle}$, where discounting method $m_j$ and order $n$ generate the minimum perplexity, will be composed with the most recent ``window" of  ${\cal S}$ on every addition to ${\cal S}$. To identify relevant terms, we use the top $m$ ranked paths of the WFST resulted from LM$_ {\langle m_j, n\rangle} \circ$ LM$_s$. This process involves a straightforward implementation of the $n$-best strings problem \cite{Nbest}, which is a generalization of the Dijkstra algorithm \cite{Dij}.  

In order to keep the number of relevant terms manageable by the analyst (as it might need to be changed based on the size of the $s$ ``window''), we used a parametric value for this step which can be changed during the application of the method (i.e. $z$ parameter in Section~\ref{sec:Formula}). After building the language models, the intersection (i.e. composition) between the two language models returns the ``relevant'' terms which can be used to measure the lexical association between $s_i$ and ${\cal D}$ and fetch those parts of $d_i$ that contain those relevant terms. Figure \ref{fig:perpel}b shows an example of the output of the 5-shortest paths. Moreover, we defined two ways by which ELICA identifies the number of relevant snippets: 
 (i) {\em Automatic- Intention Data:} By default, ELICA returns the most relevant snippet inside $d_i$, namely the one with the highest degree of lexical association with $s_i$. If the score for an identified tone ${\cal T}$ (in the range $0.5-1$) is greater than 0.75, there is a high probability that this tone indicates the perceived tone of the content of $s_i$ and it can impact the number of returned snippets\footnote{See https://console.bluemix.net/docs/services/tone-analyzer/using-tone.html\#using-the-general-purpose-endpoint for more details about interpreting tone scores.}. For example, if {\em ${\cal T}_{\text{\em tentative}}\geq$ 0.75} or {\em ${\cal T}_{\text{\em confident}}<$ 0.75}, there might be some uncertainty about the ongoing discussion in the elicitation meeting. In this case, we return {the top three} relevant snippets. On the other hand, if {\em ${\cal T}_{\text{\em confident}}$} and {\em ${\cal T}_{\text{\em analytical}} \geq$ 0.75}, ELICA presents only one (the most relevant) snippet, (ii) {\em Manual- Analysts' Choice:} {The top five} extracted relevant snippets will be presented to the analyst in the form of a list, which enables the analyst to select according to their judgment.

\section{ELICA Prototype: An Illustrative Case Study}
\label{sec:Features}
\begin{figure}
 \vspace{-3mm}
\centering
\includegraphics[scale=0.47]{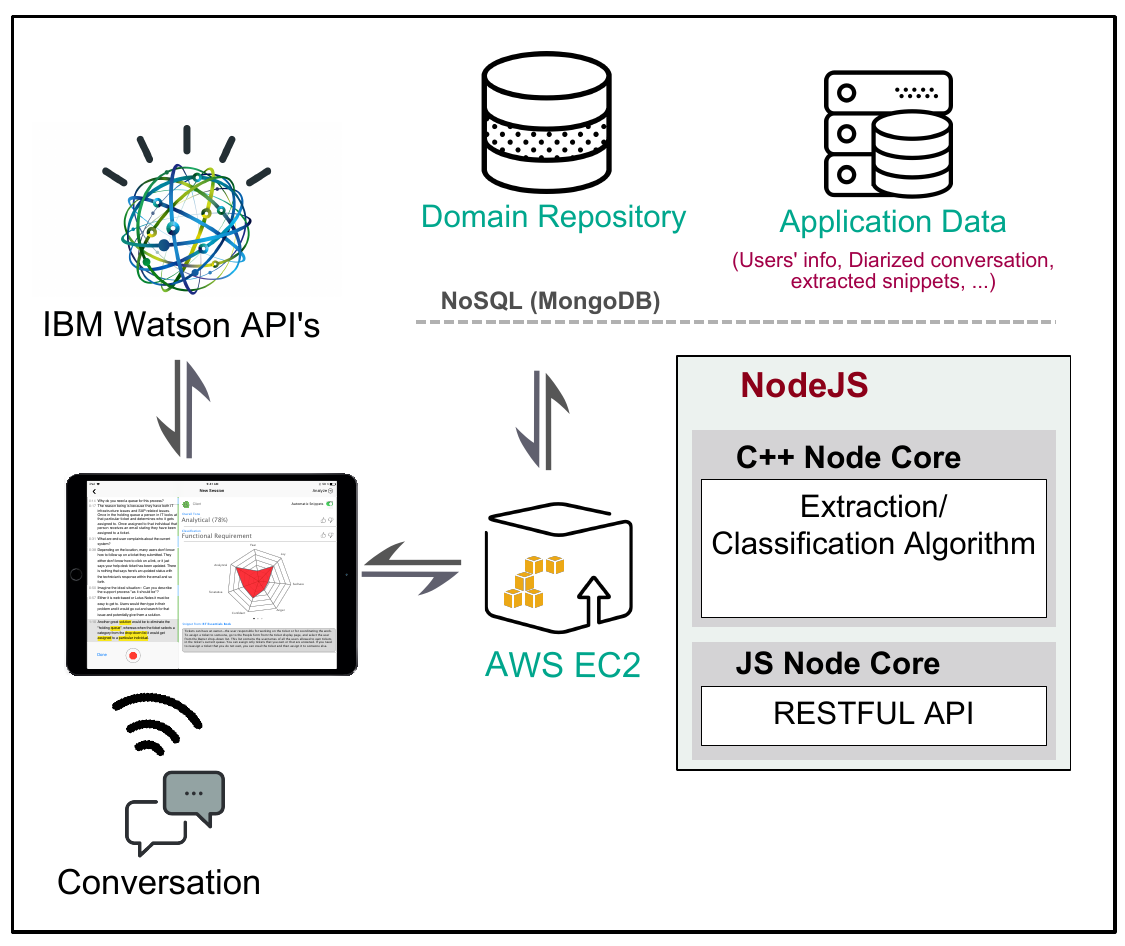}
\vspace{-3mm}
\caption{An overall architecture of ELICA. The front-end consists of an iOS app and the back-end is composed of a third party API (IBM Watson) and a single EC2 instance on Amazon Web Services (AWS). The EC2 instance contains a MongoDB database and runs node.js to expose a RESTful API.}
\label{fig:Architecture}
\vspace{-5mm}
\end{figure}

To provide an automated tool-support for the process of dynamic extraction of requirements-relevant information, we developed the ELICA prototype as a mobile app for assisting analysts during the elicitation process. Our tool is written in Swift and supports iOS version 10 and above. ELICA can be installed and used on both mobile phones and tablets. Fig.~\ref{fig:Architecture} illustrates the architecture of ELICA prototype.

In this section, we present our prototype in detail and we discuss how it realizes each of the tasks described in Section \ref{sec:App} through an illustrative case study. 
Our case study mimics the second scenario in Section \ref{sec:Formula} (Scenario \#2 in Fig.~\ref{fig:Process}) wherein an analyst is in a conversation with a client to gather the requirements of a ticketing system. The dataset for this study was provided by ThyssenKrupp Presta Steering Group USA and contains 18 transcribed interview questions obtained from two separate interviews (of 60 minutes each). The answer to each question can map to one or more requirements (32 functional and 14 non-functional requirements in total). We obtained the domain repository (${\cal D}$) for this experiment from the full source of a textbook on ticketing systems, RT Essentials \cite{RT}. RT is a high-level open-source ticketing system. The book is large enough to be used as the domain document and its subject is representative of the technical domain that the analyst might need to understand.


\begin{figure*}[!h]
\centering
\includegraphics[scale=0.6]{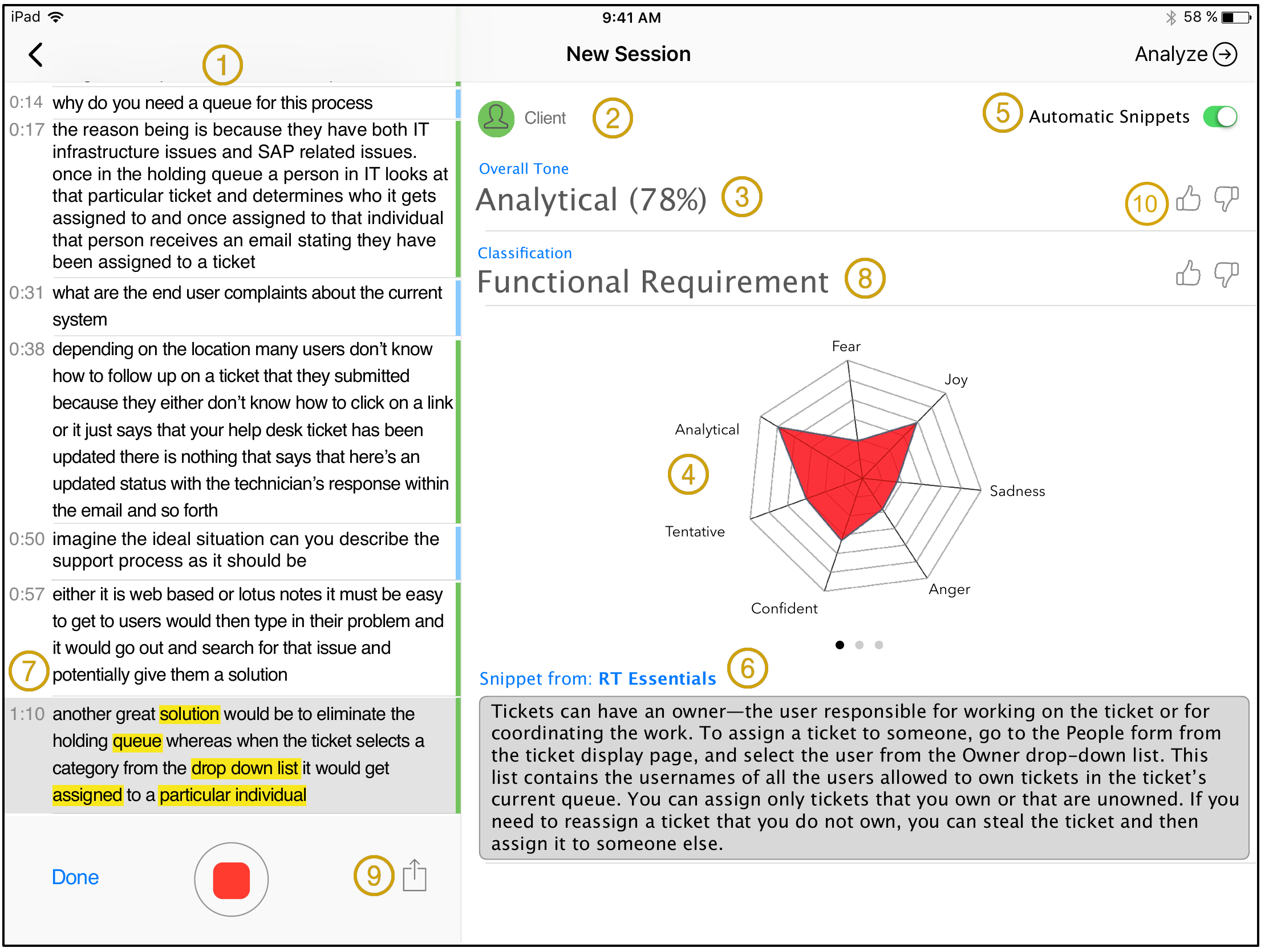}
\caption{A screenshot of the ELICA tool}
\label{fig:ipad}
\end{figure*}

\subsection{Speech to Text and Diarization}
A metadata about speakers' turns and sentence boundaries can make the transcripts more readable and can provide more context about the ongoing discussions. More specifically, in the context of requirements elicitation, knowing exactly which person is talking is important. It allows to distinguish between the client and the analyst and help to emphasize the words of the client.  
The speech-to-text and the real-time speaker diarization engines of ELICA are illustrated in Fig.~\ref{fig:ipad} (\ding{192}, \ding{193}). 

\subsection{Emotion Recognition}
Tone analysis is an NLP task to identify a tone of natural language text. In a conversation, non-verbal communication (e.g. emotions) carries important information like the intention of the speaker \cite{Emotion}. 
Therefore, understanding the text alone is not sufficient to interpret the semantics of a spoken utterance. 

To discriminate and classify the intentions that clients exhibit in either transcribed conversation (scenario \#2) or written specifications (scenario \#1), we utilize IBM Tone Analyzer API which provides an output for emotional/sentiment tone (e.g. analytical, confident, tentative, anger, cheer, and sadness).

For the purpose of data analysis (filtering the extracted snippets), we focus on only analytical, confident, and tentative emotions to identify turns containing perceived confidence or uncertainty (i.e. discussion/utterances that carry a high level of confidence/uncertainty). Other emotions detected from the text (e.g. joy, sadness, and cheer) will be used as triggers for follow-up questions during the elicitation meetings. Sections \ding{194} and \ding{195} in Fig.~\ref{fig:ipad} represent the score of the dominant emotion at the sentence-level and the distribution of emotions at the document (conversation)- level, respectively. Looking at the {\em Overall Tone} section and the Radar chart, we see that the analytical tone of the latest window of ${\cal S}$ (conversation), which is highlighted in gray, is 78\% and this tone represents the perceived tone of the full input sentence. ELICA provides two alternative visualizations  (bar chart and donut chart) for this feature which appears to be more effective for the visualization of short documents with less than three emotions as well as for the visualization of accurate data for each emotion rather than generalized relationship between emotions (the user needs to swipe right for more visualization options, using the {\ssmall \ding{108} \textcolor{Gray}{\ding{108} \ding{108}}} menu below the radar chart). 

%


\subsection{Extraction/Labeling}
The {\em extraction and classification features} (sections \ding{197}, \ding{198}, and \ding{199} in Fig.~\ref{fig:ipad}) provide a set of requirements-relevant terms (highlighted in the conversation bar) as well as the most relevant snippets from a chosen domain related document. This feature allows the analyst to not only understand the current conversation better (by highlighting the most relevant terms of $s$), but also gives them a backdrop of information to compare against and to gain more information about the context of the topic under discussion. In cases that the user activates the {\em Automatic Snippets} option (section \ding{196} in Fig.~\ref{fig:ipad}), ELICA utilizes the identified tones to filter the extracted snippets. As the most dominant tone of the most recent part of the conversation in our study is {\em analytical} with score$> 0.75$ and the tool is in the automatic mode, ELICA returns only one relevant snippet (with the highest level of lexical association). 

Regarding the domain repository ${\cal D}$, before the interview starts, the users of ELICA are able to select any set of documents to extract contextual snippets from. This allows the analyst to be able to select and choose domain repositories from the available domain documents depending on the scope of the conversation. In our case study, as illustrated in section \ding{197} of Fig.~\ref{fig:ipad}, the analyst is using the source of the RE Essentials book for this purpose. 
%
Once an interview has been completed and saved to the device, the entire session including the extracted information and intention data can then be exported to several formats such as PDF, JSON, or CSV. Also, this information can be shared with other stakeholders of the project via email, AirDrop or via any other installed mediums (section  \ding{200} in Fig.~\ref{fig:ipad}). 


\subsection{Experimental Evaluation and Results}
As discussed in Section \ref{sec:Formula}, the successful extraction of snippets from domain repository ${\cal D}$ means that the selected snippets have the highest degree of overlap with the top-ranked extracted relevant terms. To evaluate the proposed extraction method, we posed the null hypothesis: {[}$H_0${- \em contextual lexical association has no impact on the relevance of extracted snippets}{]} and used a publicly available industrial dataset (ThyssenKrupp Presta Steering Group, introduced in Section \ref{sec:Features}) to test this hypothesis. 

Given the identified snippets, to test $H_0$, we need to compute the overlap (contextual lexical association) between the reference set of relevant terms and the relevant terms included in the corresponding extracted snippets. Our reference for a correct set of relevant terms associated with each snippet was created through manual analysis by research assistants (RAs), who are experienced analysts and familiar with the problem. 
Moreover, to measure the similarity between sequences of words and more specifically between \emph{short} strings we use \emph{string edit distance} \cite{Levenshtein1}. This distance is the minimum number of edit operations (e.g. substitutions, insertions, and deletions) to transform the extracted sequence to the reference sequence. To calculate the edit distance, we use the Levenshtein edit distance \cite{Levenshtein1}.
As we could not confirm the normality of the distribution of our edit distance data, we used the non-parametric Kruskal-Wallis test to examine the null hypothesis. With p-values at 95\% significance or greater, the results of our statistical tests rejected $H_0$ at {\em p-value=0.02}, which implies the contextual lexical association can be used as an indicator of relevance in these snippets.
Moreover, our case study confirms that, on a real case with a book-sized domain document, and a lengthy real-world interview, our tool is fast enough that information can be provided in real-time, keeping pace with the conversation, and thus allowing the analyst to use the information provided by ELICA to drive the interview process.

\section{Conclusion and Future Plans}
\label{sec:Conclusion}
In this paper, we have presented a technique to dynamically extract requirements-relevant knowledge from existing documents, in order to assist analysts by surfacing relevant information from documental sources during an interactive interview. We also presented ELICA \cite{RE18}, a tool which supports the process of dynamic and automatic extraction of requirements-relevant information during elicitation meetings, in real-time. ELICA provides visual aids for its users who wish to review the extracted requirements information. 
Further, we presented an illustrative case study using an industrial dataset to simulate an elicitation meeting and to clarify various features of ELICA for supporting analysts during the elicitation process. The underlying techniques used in ELICA can also be used to extract information for other aspects of a development project, such as technical meetings with the development team, or daily stand-up/demo meetings.

As part of future work, we plan to evaluate ELICA in real software development environments. We will design studies of the usability and usefulness of our approach.  An industrial study will also allow us to assess the scalability of our approach to larger datasets and longer interviews. Moreover, as illustrated in section \ding{201}, Fig.~\ref{fig:ipad}, we record analysts' feedback on the outputs of our proposed approach. We plan to leverage this feedback to improve both extraction and intention recognition features as part of future work.
\ALT{\textcolor{applegreen}{In summary, we believe that this research will foster a clear connection between various areas of research: requirements engineering, data analysis, data visualization, Human-Computer Interaction (HCI), and cognitive psychology.}\note[vg]{This sounds a bit too imposing, for a NEXT 6-pages paper.}}

\vspace{-1mm}
\printbibliography

 
\end{document}